\documentstyle[aps,prd,preprint]{revtex}
\begin{document}
\draft
\title{Is Her X-1 a strange star?}
\author{Jes Madsen}
\address{Institute of Physics and Astronomy, University of Aarhus,
DK-8000 \AA rhus C, Denmark}
\date{November 10, 1995}
\maketitle

\begin{abstract}
The possible identification of Her X-1 with a strange star (Li et al.\
1995) is shown to be incorrect.

\end{abstract}

\vskip 2cm
\pacs{Submitted to Astronomy \& Astrophysics (Letters)}

A recent {\it Letter\/} by Li et al.\ (1995) estimates a
semiempirical mass-radius relation for the X-ray pulsar Her X-1 and
compares it with models for neutron stars and strange stars. Based on
this comparison, the authors conclude that ``the strange star model is
more consistent with Her X-1'', and therefore ``suggest it is a strange
star''. 

Unfortunately this interesting conclusion is incorrect. As demonstrated
by the authors, the concordance between the strange star
models and the Her X-1 data
occurs for a choice of bag constant $B^{1/4}$ in the range from 175--200
MeV, whereas a lower choice of bag constant gives models that are as
inconsistent with the data as are the neutron star models used. However,
strange quark matter is {\it unstable\/} for this range of parameters.
For massless quarks and negligible strong coupling (as assumed by the
authors),
stability only occurs for $B^{1/4}$ in the range from 145--164 MeV
(Farhi \& Jaffe, 1984; Madsen, 1994). This
means, that strange stars {\it cannot exist\/} for bag constants above
164 MeV, as erroneously assumed by the authors. 

Using more realistic assumptions (like finite strange quark mass and
non-zero strong coupling constant) will not improve the
situation. A non-zero strong coupling constant, $\alpha_s$,
effectively corresponds to a lowering of the bag constant and keeping
$\alpha_s=0$. This would reduce all the numbers quoted for $B^{1/4}$
above, but there would still be a gap of more than 10 MeV between the
lowest value of $B^{1/4}$ fitting Her X-1 and the highest value
consistent with strange quark matter stability. A non-zero strange quark
mass makes things even worse, because it leads to an even narrower
interval of $B^{1/4}$ for stability (and possible strange star
existence).

If the semiempirical mass-radius relation for Her X-1 derived by Li et
al.\ is correct, there is indeed an interesting problem in
interpreting it in terms of standard neutron star equations of state.
But a strange star model does not fit either.
\vfill\break

\noindent
{\bf References}

\noindent
Farhi E., Jaffe R.L., 1984, Phys.Rev.D 30, 2379

\noindent
Li X.-D., Dai Z.-G., Wang Z.-R., 1995, A\&A 303, L1

\noindent
Madsen J., 1994, Physics and Astrophysics of Strange Quark Matter. In
Sinha B., Viyogi Y.P., Raha S. (eds.) Physics and Astrophysics of
Quark-Gluon Plasma. World Scientific, Singapore, p. 186

\end{document}